\documentclass[runningheads]{llncs}
\usepackage{url}
\usepackage{hyperref}
\usepackage[backend=bibtex,style=numeric,sorting=none,url=true]{biblatex}
\usepackage[T1]{fontenc}
\usepackage{graphicx,verbatim}
\usepackage{enumitem} 
\begin{document}
\title{From Screen to Space: Evaluating Siemens’ Cinematic Reality Application for Medical Imaging on the Apple Vision Pro}
%
%
%
\titlerunning{From Screen to Space: Evaluating Siemens’ Cinematic Reality}
%
\author{Gijs Luijten\inst{1,2,3} \and
        Lisle Faray de Paiva\inst{1}\and
        Sebastian Krueger\inst{4}\and
        Alexander Brost\inst{4}\and
        Laura Mazilescu\inst{5}\and
        Ana Sofia Ferreira Santos\inst{1}\and
        Peter Hoyer\inst{6}\and
        Jens Kleesiek\inst{1,7,8,9,10,11} \and
        Sophia Marie-Therese Schmitz\inst{5}\and
        Ulf Peter Neumann\inst{5,12}\and
        Jan Egger\inst{1,2,3,8,11}}
        


\authorrunning{Luijten et al.}

\institute{Institute for Artificial Intelligence in Medicine (IKIM), Essen University Hospital (AöR), University of Duisburg-Essen, Essen, Germany \and
           Center for Virtual and Extended Reality in Medicine (ZvRM), University Hospital Essen (AöR), Essen, Germany \and
           Institute of Computer Graphics and Vision (ICG), Graz University of Technology, Graz, Austria \and
           Siemens Healthineers, Forchheim, Germany \and
           Department of General-, Visceral- and Transplant Surgery, Medical Center University Duisburg-Essen, Essen, Germany \and
           Pediatric Clinic II, University Children's Hospital Essen, University Duisburg-Essen, Essen, Germany \and
           Medical Faculty, University of Duisburg-Essen, Essen, Germany \and
           Cancer Research Center Cologne Essen (CCCE), West German Cancer Center, University Hospital Essen (AöR), Essen, Germany \and
           German Cancer Consortium (DKTK), Partner site University Hospital Essen (AöR), Essen, Germany \and
           Technische Universität Dortmund, Fakultät Physik, Dortmund, Germany\and
           Faculty of Computer Science, University of Duisburg-Essen, Essen, Germany \and
           Department of Surgery, Maastricht University Medical Centre+, Maastricht, the Netherlands\\
           \vspace{1em}
           \email{Corresponding authors: Gijs.Luijten@uk-essen.de, Jan.Egger@uk-essen.de}}

%
%
%
%
%
\maketitle              
\begin{abstract}
As one of the first research teams with full access to Siemens’ Cinematic Reality, we evaluate its usability and clinical potential for cinematic volume rendering on the Apple Vision Pro. We visualized venous-phase liver computed tomography and magnetic resonance cholangiopancreatography scans from the CHAOS and MRCP\_DLRecon datasets. Fourteen medical experts assessed usability and anticipated clinical integration potential using the System Usability Scale, ISONORM 9242-110-S questionnaire, and an open-ended survey. Their feedback identified feasibility, key usability strengths, and required features to catalyze the adaptation in real-world clinical workflows. The findings provide insights into the potential of immersive cinematic rendering in medical imaging.
\end{abstract}
\section{Introduction}
Recent advancements in extended reality (XR), encompassing augmented, mixed, and virtual reality, have greatly improved medical imaging visualization by enhancing interactivity and depth perception beyond traditional 2D displays \cite{kukla_extended_2023,lopes_explicit_2018}. Head-mounted displays (HMDs) now enable three-dimensional volume rendering (3DVR) of medical scans in immersive environments \cite{douglas_augmented_2017}. Since the late 1980s, volume rendering (VR) has facilitated 3D representations of computed tomography (CT) and magnetic resonance imaging (MRI) data on 2D monitors \cite{fuchs1989interactive}. Each voxel within an image gets assigned an opacity level and color via a so-called transfer function, creating a 3D-like image when rendered from multiple perspectives. Ongoing advancements continue expanding its medical applications \cite{ljung2016state,zhang2011volume}, with research indicating that 3DVR improves radiology diagnostics, surgical decision-making, and anatomical understanding by providing superior spatial perception compared to slice-based imaging \cite{duran2019additional,queisner2024surgical}.

Siemens’ Cinematic Reality (Siemens Healthineers, Medical Imaging Technologies, Erlangen, Germany) \cite{noauthor_cinematic_nodate} applies cinematic rendering techniques to enhance traditional VR, offering photorealistic 3D depictions using global illumination models that simulate natural light, shadows, refraction, and occlusion \cite{kroes2012exposure,ropinski2010interactive,csebfalvi2003monte,fellner2016introducing}. Prior studies demonstrated its potential, such as its application on the HoloLens (Microsoft Corp., Redmond, WA) for pediatric heart surgery planning \cite{gehrsitz2021cinematic}, but their actual benefit for surgeons in everyday clinical practice is still unclear \cite{duran2019additional}.

The Apple Vision Pro (Apple Inc., Cupertino, California, USA) offers superior resolution, computational power, and sensor capabilities compared to other HMDs, allowing for improved human-machine interaction \cite{egger2023apple}. Siemens has adapted its cinematic reality (CR) software for the Apple Vision Pro (AVP) \cite{noauthor_siemens_nodate}, warranting an investigation into its applicability beyond anatomy education, particularly for surgical planning. Liver transplantation, with its complex vascular and biliary anatomy, exemplifies a domain where precise 3D visualization could enhance clinical decision-making \cite{kelly2017depicting,yeo2018utility,erbay2003living}. Importantly, surgical planning often involves not only the identification of anatomical variants but also the assessment of tumor-induced displacement or encasement of critical structures, such as organs, vessels, and bile ducts, areas that are critical for safe and effective surgical planning, where immersive visualization could provide added clinical value.

This study evaluates Siemens’ CR on the AVP, focusing on its interactive usability and its anticipated integration into clinical workflows, such as surgical planning, as assessed through expert clinician feedback. While participants reflected on the subjective visual quality of the photorealistic rendering, these impressions were treated as part of the overall user experience rather than an isolated outcome. Although direct deployment in clinical workflows was beyond the study’s scope, the evaluation with structured usability questionnaires and domain-specific expertise provides actionable insights into the system’s readiness, features, limitations, and opportunities for clinical adoption prior to clinical trials.

Using CT and MRI, specifically magnetic resonance cholangiopancreatography (MRCP),  scans from the CHAOS \cite{kavur2021chaos,kavur_chaos_2019} and MRCP\_DLRecon \cite{kim2025deep,kim_mrcp_dlrecon_2024} datasets respectively, 14 medical experts evaluated the system via the System Usability Scale (SUS) \cite{brooke_sus_1996,SUS_Source,bangor2009determining,lewis_system_2018}, ISONORM 9242-110-S \cite{prumper_software-evaluation_1993,ISONORM_Source,bevan_iso_2015}, and qualitative questionnaires.

This study aims to contribute to a broader understanding of how immersive 3D visualization, specifically 3DVR with CR technologies on the AVP, can enhance medical imaging interpretation and support clinicians, specifically liver surgeons by identifying usability strengths, limitations, and desired features by clinicians to catalyze clinical adoption.
\begin{figure}
    \centering
        \includegraphics[width=1\textwidth]{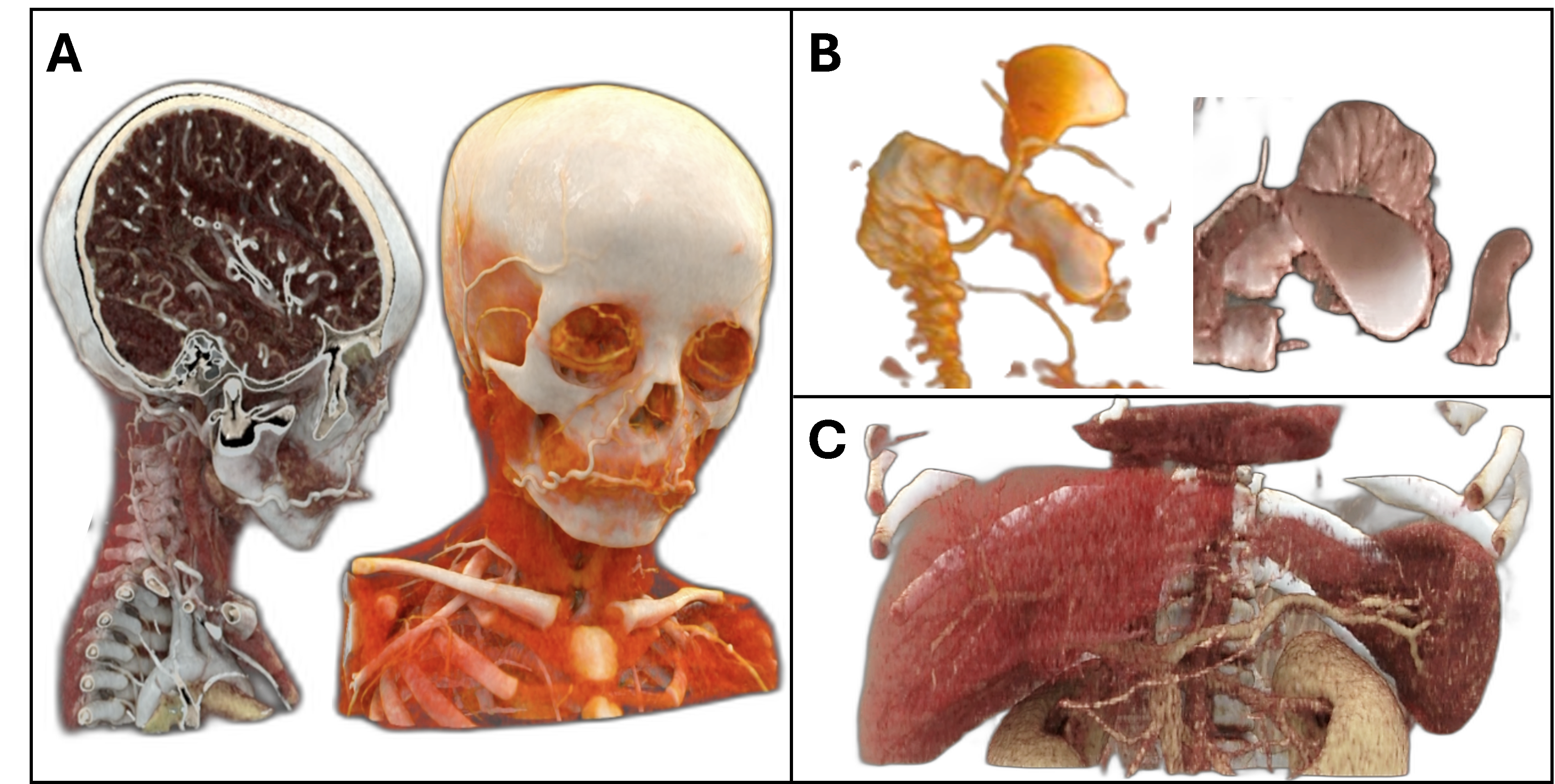}
        \caption{Cinematic 3DVR examples: Two different transfer functions and perspectives for (A) Siemens' demo case and (B) the MRCP\_DLRecon dataset case \cite{kim_mrcp_dlrecon_2024}, and (C) a single example of subject four from the CHAOS dataset \cite{kavur_chaos_2019}.}
    \label{fig:cinematicRender}
\end{figure}
\section{Methods}
\subsection{Study Design and Participants}
A mixed-methods usability evaluation was conducted, incorporating both quantitative and qualitative assessments. The quantitative analysis utilized the SUS \cite{SUS_Source} and ISONORM 9242-110-S \cite{ISONORM_Source}, which were chosen as they are validated SUS. Qualitative feedback was gathered through an open-ended survey that also included demographic information. The study participants included 11 surgeons specializing in general, visceral, vascular, and (liver) transplant surgery, one doctor assistant, and two medical students.

The entire study was conducted at the University Hospital Essen (AöR), Essen, Germany. The Ethics Committee waived the Institutional Review Board (IRB) approval requirement because the study used publicly available retrospective data and posed no risk or detriment to the participating professionals.
\subsection{Datasets}
The visualized 3D depictions on the AVP using the CR application were based on the following medical datasets:
\begin{itemize}[label=\textbullet]
    \item \textbf{CHAOS dataset} \cite{kavur2021chaos,kavur_chaos_2019}: CT volumes (512×512×Number of slices, 2 mm slice thickness) acquired during the portal venous phase after contrast agent injection, enhancing the (portal) veins and liver parenchyma.\\
    \item \textbf{MRCP\_DLRecon dataset} \cite{kim2025deep,kim_mrcp_dlrecon_2024}: An MR cholangiopancreatography (MRCP) scan (480×384×58, 1.2 mm slice thickness, 3T—3D T2w TSE sequence), selected for its detailed depiction of the biliary and pancreatic ducts, particularly relevant for hepatobiliary surgery.
\end{itemize}
From the CHAOS dataset \cite{kavur2021chaos,kavur_chaos_2019}, subjects four and five were selected to ensure a representative scan of both a male and a female, and they had 94 and 95 slices, respectively. The MRCP\_DLRecon dataset \cite{kim2025deep,kim_mrcp_dlrecon_2024} consists of a single volunteer scan with an unidentified gender. 

To prepare volumetric data for Siemens’ CR on the AVP, all scenes were generated using Siemens’ Cinematic Playground, which requires input in DICOM series format. The CHAOS dataset was already in DICOM format, while the MRCP\_DLRecon dataset, originally stored in HDF5 (.h5) format, was converted using a Python script (Python Software Foundation, Wilmington, DE, USA), utilizing the h5py, numpy, and pydicom libraries. Scenes were loaded onto the AVP via iCloud, with setup times under one minute. No manual adjustment or transfer function customization was performed; the scans were imported as-is without cropping or editing. Total time per scene preparation was thus under five minutes.
\subsection{Data Preparation for Cinematic Reality}
Siemens CR (v2.0 build 5) for the AVP (VisionOS v2.3.1) was installed via TestFlight (v3.7.1). Unlike the freely available demo version, this version supports the loading of custom-made scenes. The scenes were prepared using Siemens’ Cinematic Playground (v0.19) on a Windows laptop (Intel Core i7-11800H, NVIDIA GeForce RTX 3050) with a required Digital Imaging and Communications in Medicine (DICOM) series. The CHAOS dataset \cite{kavur_chaos_2019} provided the DICOM series directly, while the MRCP\_DLRecon dataset \cite{kim_mrcp_dlrecon_2024}, stored in Hierarchical Data Format 5 (HDF5), was converted to a DICOM series. 

The voxel values were normalized and linearly scaled [-1000 to 2000] using Python (version 12) and the following libraries: \texttt{numpy} (v2.2.2), \texttt{h5py} (v3.12.1), \texttt{nibabel} (v5.3.2), \texttt{pydicom} (v3.0.1), and \texttt{scikit-image} (v0.25.1). This scaling adjustment optimized compatibility with Siemens Cinematic Playground and Cinematic Reality’s transfer function. Scene files were then loaded onto the AVP via iCloud.

Siemens CR facilitates conventional interactions, including windowing, scrol\-ling, and zooming. The system is also capable of 3D cinematic volume rendering, as illustrated in Figure \ref{fig:cinematicRender}.
\subsection{Evaluation}
Quantitative data from the SUS \cite{SUS_Source} and ISONORM \cite{ISONORM_Source} questionnaires were analyzed using descriptive statistics, including mean scores and standard deviations. Qualitative responses were thematically categorized based on strengths, weaknesses, current clinical applicability, feature requests, and potential future applications.
The ISONORM 9241-110-S questionnaire assesses usability across seven principles derived from ISO 9241-110:
\begin{enumerate}[label=\roman*.]
    \item Suitability for the task: How well the system supports users in completing tasks.
    \item Self-descriptiveness: The intuitiveness of system functionality.
    \item Conformity with user expectations: Consistency with known interface standards.
    \item Learnability: Ease with which new users can become proficient.
    \item Controllability: The extent to which users can influence actions and outcomes.
    \item Error tolerance: The system’s ability to prevent or recover from user errors.
    \item Customizability: How well the system can be adapted to user needs.
\end{enumerate}
These principles provide a comprehensive framework for evaluating the user-system interaction, focusing on effectiveness, efficiency, and satisfaction in a specified context of use \cite{ISO9241_110}.

The System Usability Scale (SUS) is a widely used tool for evaluating the usability of various systems, including medical software. It consists of a 10-item questionnaire with five response options for respondents, ranging from strongly agree to strongly disagree. SUS provides a single score ranging from 0 to 100, representing a composite measure of the overall usability of the system being studied \cite{SUS_Source}. Notably, scores above 68 are considered above average across industries and for digital health applications, with higher scores indicating better usability \cite{hyzy_system_2022,10.5555/2817912.2817913}. Scores between 74 and 85 indicate good to excellent usability \cite{10.5555/2817912.2817913}.
\subsection{Procedure and Data Collection}
\begin{figure}
    \centering
        \includegraphics[width=1\textwidth]{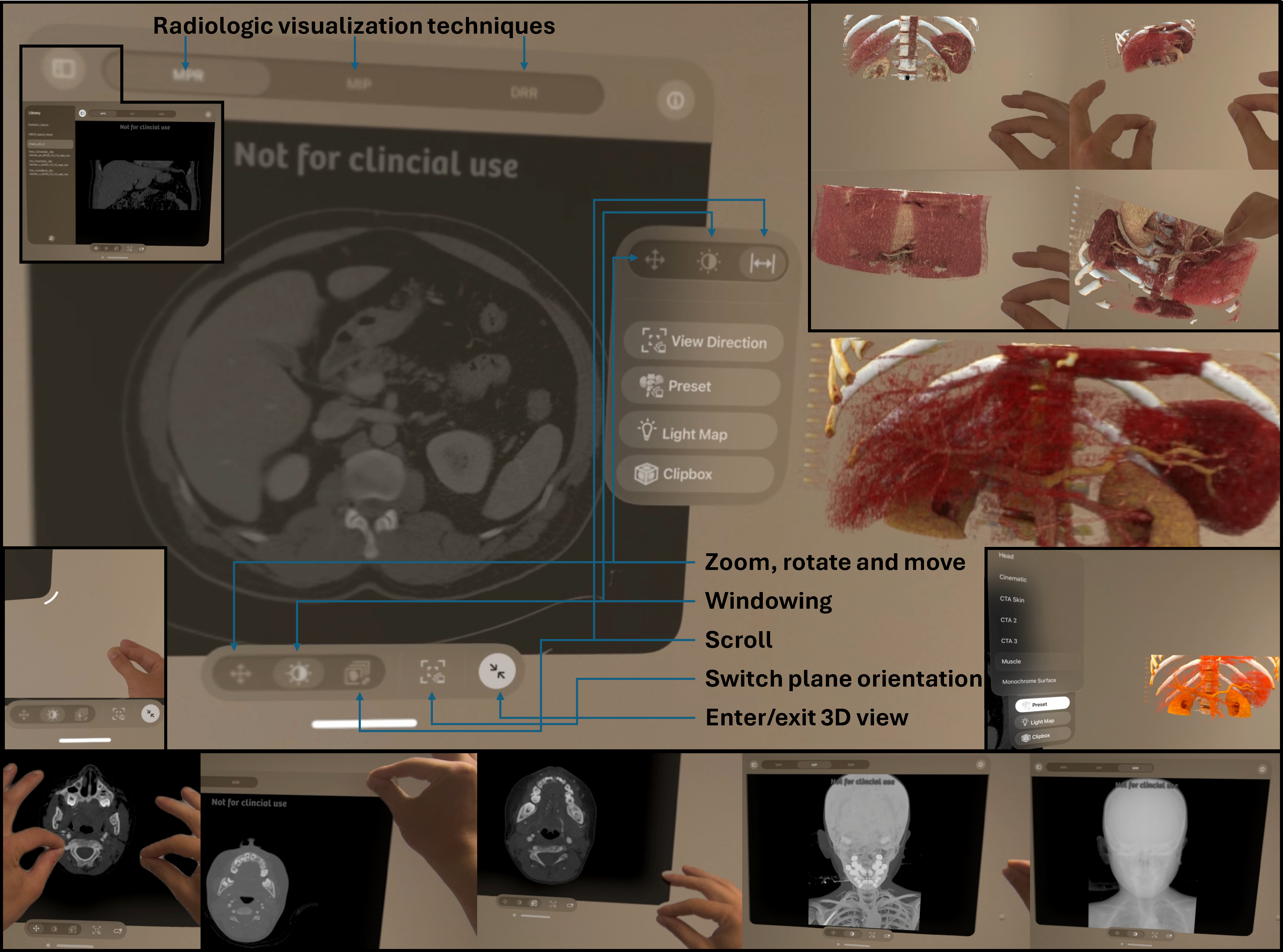}
        \caption{Core functionalities of Siemens’ Cinematic Reality on the Apple Vision Pro.
Eye gaze functions as a pointer, and finger pinching acts as a selection (click) tool.
Top left: Access the library to load scenes previously generated from DICOM data.
Top right: 3D model interactions include scrolling/clipping (top left), resizing (top right), windowing (bottom left), and rotation (bottom right).
Presets enable tissue-specific transfer functions (e.g., orange rendering in the lower-right model); clipping planes/boxes and light map settings adjust visibility and lighting effects.
View orientation can be switched via the 2D/3D overview (axial, sagittal, coronal).
Bottom: From left to right, 2D slice interactions include zooming (two hands), windowing, scrolling, and switching between radiologic visualization modes.
Rotation supports one or two hands; all other interactions use a single hand.}
    \label{fig:CRoverviewUI}
\end{figure}

\begin{itemize}[label=\textbullet]
    \item \textbf{Informed Consent:} All participants provided written informed consent prior to study participation.
\end{itemize}
\begin{itemize}[label=\textbullet]
    \item \textbf{Device Setup:}
        \begin{itemize}[label=\textbullet]
            \item Streaming from the AVP to an iPad was activated via AirPlay to facilitate observation and assistance.
            \item Participants put on the AVP headset.
            \item Each participant completed the built-in hand-eye calibration before use of Siemens' CR application. The researcher guided participants in this process.
        \end{itemize}
\end{itemize}
\begin{itemize}[label=\textbullet]
    \item \textbf{Application Interaction (Task):}
    \begin{itemize}[label=\textbullet]
        \item Participants launched the Siemens Cinematic Reality (CR) application, which was preloaded with CR scenes from the CHAOS and MRCP\_DLRecon datasets. The datasets were presented in a fixed order: participants first explored the CT volume from the CHAOS dataset, followed by the MRCP scan. This sequence enabled them to examine both vascular and biliary anatomy and assess the utility of CR in distinct clinical contexts. After viewing both datasets, participants were free to revisit scenes or explore an additional demo head CT case provided by Siemens. From this point onward, participants were not guided; help was only provided upon request from the participant. 
        
        Participants were tasked with the following instruction:
        \begin{itemize}[label=\textbullet]
            \item Explore key anatomical structures relevant to hepatobiliary surgery, including portal veins, bile ducts, and liver parenchyma.
            \item Examine image fidelity and photorealism using cinematic volume rendering.
            \item Navigate the interface, Figure \ref{fig:CRoverviewUI}, using standard interaction techniques such as zooming, panning, scrolling, windowing, and rotating the volume and traditional CT slices. Interact with the volume by slicing it, adapting the transfer function, and switching between scenes (Siemens' demo scene, CHAOS, MRCP\_DLRecon).
            \item Verbally articulate their observations, usability impressions, and any encountered challenges as part of a “think-aloud” protocol. 
            \item Participants were asked to describe how they would ideally integrate the system into their clinical workflow, for example in surgical planning or patient case discussions, and to articulate which functionalities they would use and how. These interactions and comments were observed by the researcher in real time via AirPlay streaming to an iPad.
        \end{itemize}
        \item If assistance was required, the researcher provided guidance, utilizing the AirPlay stream to facilitate real-time support.
    \end{itemize}
\end{itemize}
While no rigid task list or time limit was imposed, the average interaction time during the application interaction (task) was approximately 15–20 minutes, while hand-eye calibration took less than five minutes. Researchers provided minimal assistance, only intervening when participants requested support, facilitated through the mirrored iPad view.\\

\noindent\textbf{Data Collection:}\\
Following the interaction session, participants completed the System Usability Scale (SUS), the ISONORM 9242-110-S questionnaire, and an open-ended survey capturing qualitative feedback. All responses were collected via Google Forms.

%
%
%
%
%
\section{Results and Discussion}
\subsection{Participant Demographics}
Fourteen subjects participated in the study, including two medical students, a doctor/surgeon's assistant, and 11 surgeons, with a gender distribution of eight females and six males. The ages of the female subjects ranged from 22 to 43 years, with a mean of $\mu=32.0$ years, standard deviation $\sigma=6.70$, and a median of 32 years $(IQR=7.25; Q_1=28.75$, $Q_3=36.00)$. In contrast, males, ranging in age from 35 to 68 years, exhibited a mean age of $\mu=48.83$ years, standard deviation $\sigma=11.92$, and a median of 46 years $(IQR=13.00; Q_1=41.75$, $Q_3=54.75)$.

The AVP does not permit the use of spectacles within the HMD; rather, it necessitates the acquisition of its proprietary insert lenses. Consequently, the diopter of the subjects is an essential factor. Among the ten participants who wear spectacles, one subject was classified as farsighted $(+2.0)$, while the remaining subjects were nearsighted $(\mu=-3.22$, $\sigma=2.91$, minimum $-9.25$, maximum $-1.0)$. 
Of the ten participants who normally wear spectacles, eight used the AVP with compatible insert lenses. The remaining two completed the study without visual correction. While both were able to complete all tasks, they reported reduced clarity when inspecting fine anatomical structures. Although this did not prevent participation, it may have influenced their subjective usability ratings, exemplifying the need for vision correction support in clinical extended reality systems.

Notably, only two subjects had prior experience with HMDs, which was limited to one and eight hours, respectively. The HMDs were the HoloLens 2 and/or Meta Quest 3 (Meta Platforms, Inc., Menlo Park, CA). All subjects were unfamiliar with 3DVR; their familiarity with 3D rendering was limited to segmentation-based 3D rendering.
\subsection{System Usability Scale}
The SUS is technology-agnostic and combines effectiveness, efficiency and satisfaction with high reliability (Cronbach alpha = 0.91) in a single score (0-100) \cite{bangor2009determining,bangor2008empirical,vlachogianni2022perceived,SUS_Source}. Based on over ten years of empirical evidence, the score can be divided into seven categories, from worst to best imaginable  \cite{bangor2009determining,vlachogianni2022perceived}.

The mean score was $\sigma=77.68$ $(\sigma=15.01, IQR=27.50; Q_1=63.75, Q_3=91.25)$, indicating a score between good and excellent. Surgeons and assistant doctors rated the system higher than students. There were no noticeable differences between age or gender groups. An overview of the results for each group is shown in Figure 2, including students, residents, and surgeons.
\begin{figure}
    \centering
    \includegraphics[width=1\textwidth]{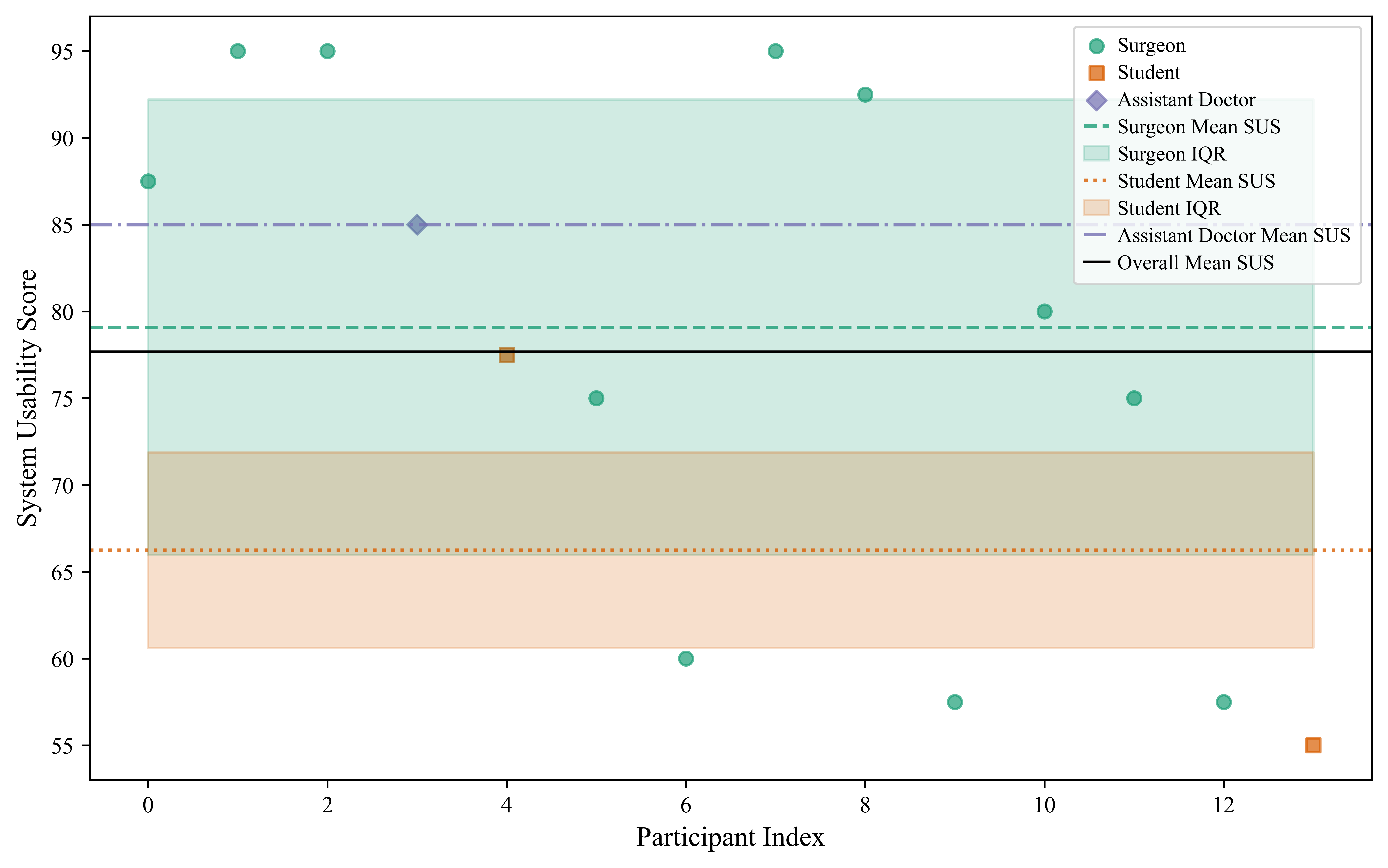}
    \caption{Visualization of SUS scores for each participant, grouped by profession. Mean SUS scores are indicated by horizontal lines, with shaded areas representing the Interquartile Range (IQR).}
    \label{fig:sus_scores}
\end{figure}
\newpage
\subsection{ISONORM 9242-110-S}
The ISONORM questions can be grouped into seven different measurements: suitability $(\mu =5.76, \sigma=0.55, \tilde{x}=6, IQR=0.67; Q_1=5.33, Q_3=6)$, self-descriptiveness $(\mu=5.21, \sigma=1.13, \tilde{x}=5.17, IQR=1.33; Q_1=4.67, Q_3=6)$, conformity $(\mu=5.74, \sigma=0.71, \tilde{x}=6, IQR=0.83; Q_1=5.17, Q_3=6)$, learnability $(\mu 5.5, \sigma=0.88, \tilde{x}=5.67, IQR = 1; Q_1=5, Q_3=6)$, controllability $(\mu=5.81, \sigma=0.77, \tilde{x}=6, IQR=1; Q_1=5.33, Q_3=6.33)$, error tolerance $(\mu=5.24, \sigma=1.02, \tilde{x}=5.17, IQR=1.5; Q_1=4.50, Q_3=6.00)$, and customizability $(\mu=5.19, \sigma=1.19, \tilde{x}=5, IQR=1.25; Q_1=4.75, Q_3=6)$.

\begin{figure}
    \centering
    \includegraphics[width=\textwidth]{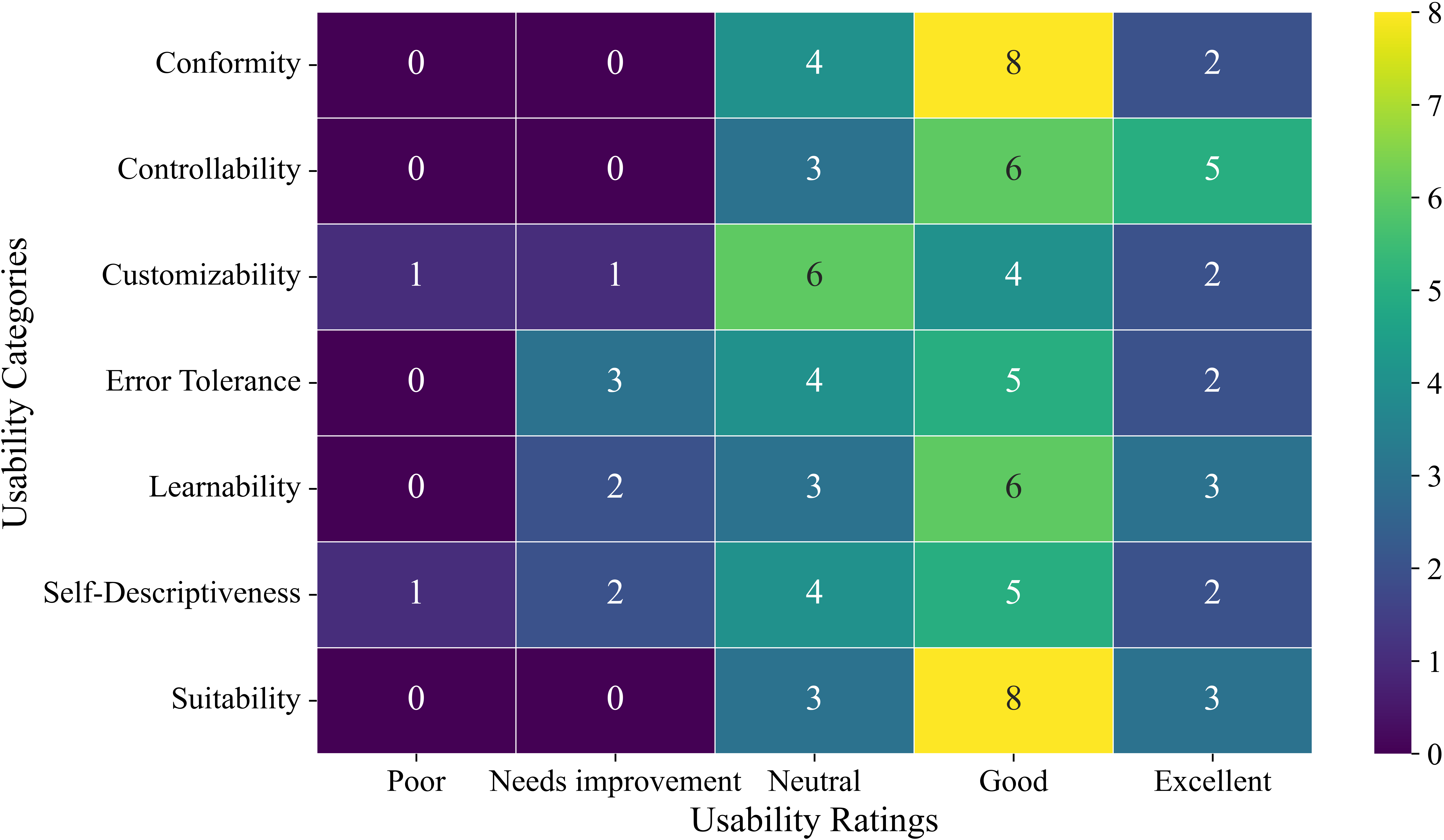}
    \caption{Heatmap of ISONORM responses grouped per usability measurement.}
    \label{fig:heatmap_isonorm}
\end{figure}

\begin{figure}
    \centering
    \includegraphics[width=1\textwidth]{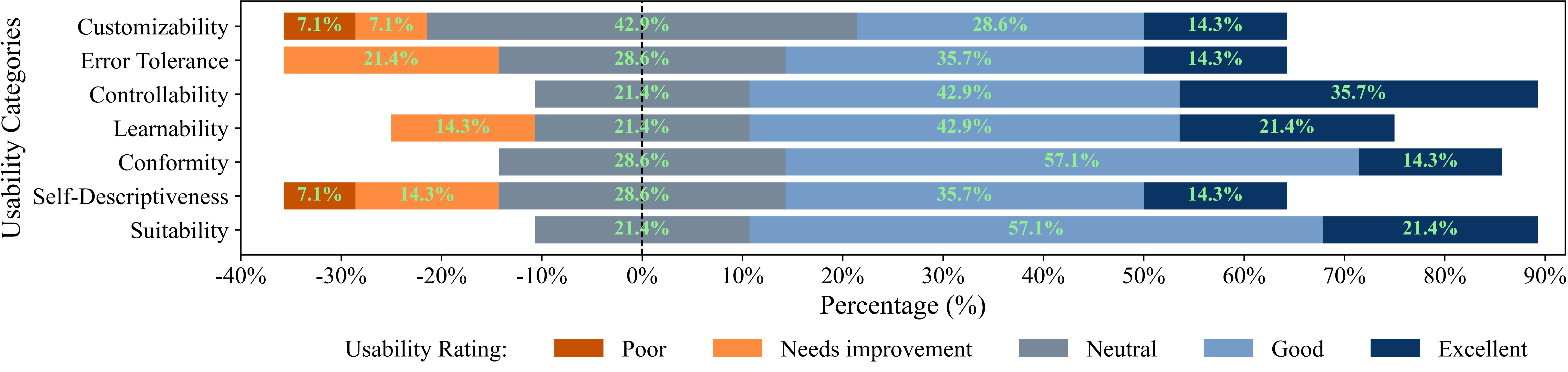}
    \caption{Stacked bar plot of ISONORM responses grouped per usability measurement centered around 0\%.}
    \label{fig:stacked_bar_plot_isornorm}
\end{figure}

The heat map and stacked bar plot shown in Figures \ref{fig:heatmap_isonorm} and \ref{fig:stacked_bar_plot_isornorm} show high conformity and suitability, with no need for improvement in controllability. In contrast, self-descriptiveness and customizability had the most room for improvement. 
\subsection{Qualitative Feedback}
The open-ended questionnaire delved into the advantages and disadvantages of the CR application on the AVP (Table \ref{tab:positive_negative}), its potential applications in a medical setting and its current clinical usability (Table \ref{tab:medical_use_cases}), as well as specific complex use cases and desired improvements for both software.

Surgeons identified key areas for enhancing the Siemens CR application on AVP to optimize its clinical utility. Recommended improvements include integration with electronic patient records, advanced measurement tools for tumor and vessel size assessment, and the ability to insert 3D models of surgical instruments. Enhanced segmentation functionality enabling seamless switching between anatomical structures (e.g., arteries, portal veins, bile ducts), while gesture-based removal of irrelevant areas should provide more focused visualizations. Additionally, multi-user and multi-platform compatibility to facilitate collaborative planning and real-time surgical discussions. Finally, incorporating a voice activation function was suggested to enhance operability in sterile settings or when both hands are occupied with other tasks.

The setup process for AVP was typically less than five minutes and considered straightforward. This duration includes donning the headset, performing the initial eye and hand calibration, and launching the Cinematic Reality (CR) application. Such a setup time is likely more acceptable for preoperative planning sessions, where time constraints are less critical. However, this setup duration may be undesirable for intraoperative applications, but the participants did not see this as a problem.
\begin{table}
\centering
\caption{Positive and negative aspects regarding Siemens' Cinematic Reality app of the Apple Vision Pro.}
\label{tab:positive_negative}
{\fontsize{9pt}{12pt}\selectfont
\begin{tabular}{|p{0.494\textwidth}|p{0.494\textwidth}|} 
\hline
\textbf{Positive aspects} & \textbf{Negative aspects} \\
\hline
Intuitive and easy to use & Lack of segmentation options \\
\hline
Exceptional user experience & Lack of annotation options \\
\hline
\parbox[t]{6cm}{Accurate 3D visualization from\\all angles} & \parbox[t]{6cm}{ Lack of drawing 2D/3D lines and \\ measurement} \\
\hline
High-resolution 3D models and images & \parbox[t]{6cm}{Incomplete 3D reconstruction \\(e.g., missing pathologies) \\Note: also not visible in the 2D slices\vspace{0.8mm}} \\
\hline
Superior 2D display resolution and comfort & \parbox[t]{6cm}{No fusion options or multi-scan \\ visualization\vspace{0.8mm}} \\
\hline
Instant patient-specific 3D reconstruction & Unclear menu for the 3D navigation \\
\hline
Efficient/high-quality zoom functionality & Incompatibility with personal eyewear \\
\hline
Adaptive screen and model positioning & Mild nausea after use \\
\hline
Interaction via eye-tracking & Prolonged scene loading time (<5 seconds) \\
\hline
Seamless synchronization and streaming &  \\
\hline
\end{tabular}
}
\end{table}
\begin{table}
\centering
\caption{Potential medical use cases and feasibility in daily clinical routine for the Apple Vision Pro.}
\label{tab:medical_use_cases}
{\fontsize{9pt}{12pt}\selectfont
\begin{tabular}{|p{0.19\textwidth}|p{0.4\textwidth}|p{0.4\textwidth}|} 
\hline
\textbf{Medical Application} & \textbf{Potential Medical Use Cases} & \textbf{Feasibility in Clinical Daily Routine} \\
\hline
Education \& Training & Medical education (e.g., ultrasound-anatomy correlation), patient information, student learning & Useful for student education and patient information \\
\hline
Preoperative Planning & Surgical planning, anatomical assessment (e.g., tumor/vessel measurement, anatomical variants) & Feasible for complex surgical planning, vascular anatomy assessment \\
\hline
Intraoperative Guidance & Overlay on organs, adjusting surgery to anatomical variations & Enhances visualization of complex anatomy for improved surgical steps \\
\hline
Diagnostic \& Imaging Support & Pathology localization, radiology & Could supplement CT imaging for spatial understanding \\
\hline
Intervention \& Procedures & Puncture guidance, intervention planning & Hard to assess without a clinical study in the operating room \\
\hline
Surgical Discussions & Surgical planning discussions, case presentations & Case discussions and general surgeons' preparation \\
\hline
Operations & Direct application in surgery (e.g., overlay, quick consultant, entry path, resection planes) & Unclear feasibility; some users see potential, others do not. \\
\hline
\end{tabular}
}
\end{table}
\newpage
\section{Conclusion}
According to the participating surgeons, this study confirms the potential of immersive 3D cinematic 3DVR on the AVP to enhance medical imaging interpretation.

The findings indicate high usability ratings and an intuitive user interface that facilitates detailed, photorealistic CT and MRI data visualization. The interface is particularly well-suited for education, surgical planning, and intraoperatively recalling 3D spatial anatomical relations, especially in complex cases and with anatomical variations. 

It is noteworthy that the open-ended feedback highlighted valuable improvements for clinical adoption, such as enhanced segmentation, advanced measurement and annotation tools, and smoother integration with clinical systems. Moreover, this work invites further investigation to ascertain the extent to which surgical planning will be influenced by the utilization of this software on the AVP during both the pre- and intraoperative phases.

\section{Acknowledgements}
Ana Sofia Ferreira Santos receives funding by the European Union under Grant Agreement 101168715. Views and opinions expressed are however those of the author(s) only and do not necessarily reflect those of the European Union. Neither the European Union nor the granting authority can be held responsible for them.
This work was supported by the REACT-EU project KITE (grant number: EFRE-0801977, Plattform für KI-Translation Essen, https://kite.ikim.nrw/).
%
%
%
%
%
\printbibliography

@article{kukla_extended_2023,
	title = {Extended {Reality} in {Diagnostic} {Imaging}—{A} {Literature} {Review}},
	volume = {9},
	issn = {2379-1381},
	doi = {10.3390/tomography9030088},
	number = {3},
	urldate = {2025-02-17},
	journal = {Tomography},
	author = {Kukla, Paulina and Maciejewska, Karolina and Strojna, Iga and Zapał, Małgorzata and Zwierzchowski, Grzegorz and Bąk, Bartosz},
	month = may,
	year = {2023},
	pmid = {37368540},
	pmcid = {PMC10303875},
	pages = {1071--1082},
}

@article{douglas_augmented_2017,
	title = {Augmented {Reality}: {Advances} in {Diagnostic} {Imaging}},
	volume = {1},
	copyright = {http://creativecommons.org/licenses/by/3.0/},
	issn = {2414-4088},
	doi = {10.3390/mti1040029},
	language = {en},
	number = {4},
	urldate = {2025-02-17},
	journal = {Multimodal Technologies and Interaction},
	author = {Douglas, David B. and Wilke, Clifford A. and Gibson, J. David and Boone, John M. and Wintermark, Max},
	month = dec,
	year = {2017},
	note = {Number: 4
Publisher: Multidisciplinary Digital Publishing Institute},
	pages = {29},
}

@article{fuchs1989interactive,
  title={Interactive visualization of 3D medical data},
  author={Fuchs, Henry and Levoy, Marc and Pizer, Stephen M},
  journal={Computer},
  volume={22},
  number={8},
  year={1989}
}

@inproceedings{ljung2016state,
  title={State of the art in transfer functions for direct volume rendering},
  author={Ljung, Patric and Kr{\"u}ger, Jens and Groller, Eduard and Hadwiger, Markus and Hansen, Charles D and Ynnerman, Anders},
  booktitle={Computer graphics forum},
  volume={35},
  number={3},
  pages={669--691},
  year={2016},
  organization={Wiley Online Library}
}

@article{zhang2011volume,
  title={Volume visualization: a technical overview with a focus on medical applications},
  author={Zhang, Qi and Eagleson, Roy and Peters, Terry M},
  journal={Journal of digital imaging},
  volume={24},
  pages={640--664},
  year={2011},
  publisher={Springer}
}

@inproceedings{csebfalvi2003monte,
  title={Monte carlo volume rendering},
  author={Cs{\'e}bfalvi, Bal{\'a}zs and Szirmay-Kalos, L{\'a}szl{\'o}},
  booktitle={IEEE Visualization, 2003. VIS 2003.},
  pages={449--456},
  year={2003},
  organization={IEEE}
}

@inproceedings{ropinski2010interactive,
  title={Interactive volumetric lighting simulating scattering and shadowing},
  author={Ropinski, Timo and D{\"o}ring, Christian and Rezk-Salama, Christof},
  booktitle={2010 ieee pacific visualization symposium (pacificvis)},
  pages={169--176},
  year={2010},
  organization={IEEE}
}

@article{kroes2012exposure,
  title={Exposure render: An interactive photo-realistic volume rendering framework},
  author={Kroes, Thomas and Post, Frits H and Botha, Charl P},
  journal={PloS one},
  volume={7},
  number={7},
  pages={e38586},
  year={2012},
  publisher={Public Library of Science San Francisco, USA}
}

@article{duran2019additional,
  title={The additional diagnostic value of the three-dimensional volume rendering imaging in routine radiology practice},
  author={Duran, Alper H and Duran, Munevver N and Masood, Irfan and Maciolek, Lynsey M and Hussain, Huda},
  journal={Cureus},
  volume={11},
  number={9},
  year={2019},
  publisher={Cureus Inc.}
}

@article{queisner2024surgical,
  title={Surgical planning in virtual reality: a systematic review},
  author={Queisner, Moritz and Eisentr{\"a}ger, Karl},
  journal={Journal of Medical Imaging},
  volume={11},
  number={6},
  pages={062603--062603},
  year={2024},
  publisher={Society of Photo-Optical Instrumentation Engineers}
}

@article{gehrsitz2021cinematic,
  title={Cinematic rendering in mixed-reality holograms: a new 3D preoperative planning tool in pediatric heart surgery},
  author={Gehrsitz, Pia and Rompel, Oliver and Sch{\"o}ber, Martin and Cesnjevar, Robert and Purbojo, Ariawan and Uder, Michael and Dittrich, Sven and Alkassar, Muhannad},
  journal={Frontiers in Cardiovascular Medicine},
  volume={8},
  pages={633611},
  year={2021},
  publisher={Frontiers Media SA}
}

@article{fellner2016introducing,
  title={Introducing cinematic rendering: a novel technique for post-processing medical imaging data},
  author={Fellner, Franz A},
  journal={Journal of Biomedical Science and Engineering},
  volume={9},
  number={3},
  pages={170--175},
  year={2016},
  publisher={Scientific Research Publishing}
}

@article{egger2023apple,
	title = {Is the {Apple} {Vision} {Pro} the {Ultimate} {Display}? {A} {First} {Perspective} and {Survey} on {Entering} the {Wonderland} of {Precision} {Medicine}},
	volume = {12},
	doi = {10.2196/52785},
	number = {1},
	urldate = {2025-02-21},
	journal = {JMIR Serious Games},
	author = {Egger, Jan and Gsaxner, Christina and Luijten, Gijs and Chen, Jianxu and Chen, Xiaojun and Bian, Jiang and Kleesiek, Jens and Puladi, Behrus},
	month = sep,
	year = {2024},
	note = {Company: JMIR Serious Games
Distributor: JMIR Serious Games
Institution: JMIR Serious Games
Label: JMIR Serious Games
Publisher: JMIR Publications Inc., Toronto, Canada},
	pages = {e52785},
}

@misc{noauthor_siemens_nodate,
  title        = {Siemens Healthineers launches Cinematic Reality app for Apple Vision Pro},
  url = {https://www.siemens-healthineers.com/press/releases/cinematic-reality-applevisionpro},
  note         = {last Accessed: 2025/02/17}
}

@article{kelly2017depicting,
  title={Depicting surgical anatomy of the porta hepatis in living donor liver transplantation},
  author={Kelly, Paul and Fung, Albert and Qu, Joy and Greig, Paul and Tait, Gordon and Jenkinson, Jodie and McGilvray, Ian and Agur, Anne},
  journal={Journal of visualized surgery},
  volume={3},
  year={2017},
  publisher={AME Publications}
}

@article{yeo2018utility,
  title={Utility of 3D reconstruction of 2D liver computed tomography/magnetic resonance images as a surgical planning tool for residents in liver resection surgery},
  author={Yeo, Caitlin T and MacDonald, Andrew and Ungi, Tamas and Lasso, Andras and Jalink, Diederick and Zevin, Boris and Fichtinger, Gabor and Nanji, Sulaiman},
  journal={Journal of surgical education},
  volume={75},
  number={3},
  pages={792--797},
  year={2018},
  publisher={Elsevier}
}

@article{erbay2003living,
  title={Living donor liver transplantation in adults: vascular variants important in surgical planning for donors and recipients},
  author={Erbay, Nazli and Raptopoulos, Vassilios and Pomfret, Elizabeth A and Kamel, Ihab R and Kruskal, Jonathan B},
  journal={American Journal of Roentgenology},
  volume={181},
  number={1},
  pages={109--114},
  year={2003},
  publisher={Am Roentgen Ray Soc}
}

@article{kavur2021chaos,
	title = {{CHAOS} {Challenge} - combined ({CT}-{MR}) healthy abdominal organ segmentation},
	volume = {69},
	issn = {1361-8415},
	doi = {10.1016/j.media.2020.101950},
	urldate = {2025-02-21},
	journal = {Medical Image Analysis},
	author = {Kavur, A. Emre and Gezer, N. Sinem and Barış, Mustafa and Aslan, Sinem and Conze, Pierre-Henri and Groza, Vladimir and Pham, Duc Duy and Chatterjee, Soumick and Ernst, Philipp and Özkan, Savaş and Baydar, Bora and Lachinov, Dmitry and Han, Shuo and Pauli, Josef and Isensee, Fabian and Perkonigg, Matthias and Sathish, Rachana and Rajan, Ronnie and Sheet, Debdoot and Dovletov, Gurbandurdy and Speck, Oliver and Nürnberger, Andreas and Maier-Hein, Klaus H. and Bozdağı Akar, Gözde and Ünal, Gözde and Dicle, Oğuz and Selver, M. Alper},
	month = apr,
	year = {2021},
	pages = {101950},
}

@article{kim2025deep,
  title={Deep Learning-Based Accelerated MR Cholangiopancreatography Without Fully-Sampled Data},
  author={Kim, Jinho and Nickel, Marcel Dominik and Knoll, Florian},
  journal={NMR in Biomedicine},
  volume={38},
  number={3},
  pages={e70002},
  year={2025},
  publisher={Wiley Online Library}
}

@misc{kim_mrcp_dlrecon_2024,
    author = {Kim, Jinho and Nickel, Dominik and Knoll, Florian},
	title = {{MRCP\_DLRecon (v3) [Data set]. Zenodo}},
	url = {https://zenodo.org/records/13912092},
}

@misc{kavur_chaos_2019,
  author       = {Ali Emre Kavur and M. Alper Selver and Oğuz Dicle and Mustafa Barış and  N. Sinem Gezer},
  title        = {{CHAOS - Combined (CT-MR) Healthy Abdominal Organ Segmentation Challenge Data (v1.03) [Data set]. The IEEE International Symposium on Biomedical Imaging (ISBI), Venice, Italy. Zenodo.}},
  month        = Apr,
  year         = 2019,
  publisher    = {Zenodo},
  version      = {v1.03},
  url          = {https://doi.org/10.5281/zenodo.3362844}
}

@incollection{brooke_sus_1996,
	title = {{SUS}: {A} '{Quick} and {Dirty}' {Usability} {Scale}},
	isbn = {978-0-429-15701-1},
	shorttitle = {{SUS}},
	booktitle = {Usability {Evaluation} {In} {Industry}},
	publisher = {CRC Press},
	author = {Brooke, john},
	year = {1996},
	note = {Num Pages: 6},
}

@article{lewis_system_2018,
	title = {The {System} {Usability} {Scale}: {Past}, {Present}, and {Future}},
	volume = {34},
	issn = {1044-7318},
	shorttitle = {The {System} {Usability} {Scale}},
	doi = {10.1080/10447318.2018.1455307},
	number = {7},
	urldate = {2025-02-17},
	journal = {International Journal of Human–Computer Interaction},
	author = {Lewis, James R.},
	month = jul,
	year = {2018},
	note = {Publisher: Taylor \& Francis},
	pages = {577--590},
}

@inproceedings{bevan_iso_2015,
	address = {Cham},
	title = {{ISO} 9241-11 {Revised}: {What} {Have} {We} {Learnt} {About} {Usability} {Since} 1998?},
	isbn = {978-3-319-20901-2},
	shorttitle = {{ISO} 9241-11 {Revised}},
	doi = {10.1007/978-3-319-20901-2_13},
	booktitle = {Human-{Computer} {Interaction}: {Design} and {Evaluation}},
	publisher = {Springer International Publishing},
	author = {Bevan, Nigel and Carter, James and Harker, Susan},
	editor = {Kurosu, Masaaki},
	year = {2015},
	pages = {143--151},
}

@book{prumper_software-evaluation_1993,
	title = {Software-{Evaluation} {Based} upon {ISO} 9241 {Part} 10},
	volume = {733},
	isbn = {978-3-540-57312-8},
	author = {Prümper, Jochen},
	month = jan,
	year = {1993},
	doi = {10.1007/3-540-57312-7_74},
	note = {Pages: 265},
}

@article{bangor2009determining,
  title={Determining what individual SUS scores mean: Adding an adjective rating scale},
  author={Bangor, Aaron and Kortum, Philip and Miller, James},
  journal={Journal of usability studies},
  volume={4},
  number={3},
  pages={114--123},
  year={2009},
  publisher={Usability Professionals' Association Bloomingdale, IL}
}

@article{bangor2008empirical,
  title={An empirical evaluation of the system usability scale},
  author={Bangor, Aaron and Kortum, Philip T and Miller, James T},
  journal={Intl. Journal of Human--Computer Interaction},
  volume={24},
  number={6},
  pages={574--594},
  year={2008},
  publisher={Taylor \& Francis}
}

@article{vlachogianni2022perceived,
  title={Perceived usability evaluation of educational technology using the System Usability Scale (SUS): A systematic review},
  author={Vlachogianni, Prokopia and Tselios, Nikolaos},
  journal={Journal of Research on Technology in Education},
  volume={54},
  number={3},
  pages={392--409},
  year={2022},
  publisher={Taylor \& Francis}
}

@misc{SUS_Source,
  title        = {System Usability Scale (SUS) Questionnaire - Online source by John Brooke},
  url = {https://rickvanderzwet.nl/trac/personal/export/104/liacs/hci/docs/SUS-questionaire.pdf},
  note         = {last accessed: 2025/02/18}
}

@misc{ISONORM_Source,
  title        = {ISONORM 9241/110-S - Online source by Jochen Prümper},
  url = {http://www.uselab.tu-berlin.de/wiki/images/6/62/ISONorm\_Kurzversion.pdf},
  note         = {last accessed: 2025/02/21}
}

@misc{noauthor_cinematic_nodate,
	type = {Text},
	title = {Cinematic {Rendering} in medical imaging},
	copyright = {2025},
	url = {https://www.siemens-healthineers.com/digital-health-solutions/cinematic-rendering},
    note = {last accessed 2025/2/25}
}

@article{lopes_explicit_2018,
	title = {Explicit design of transfer functions for volume-rendered images by combining histograms, thumbnails, and sketch-based interaction},
	journal = {The Visual Computer},
	author = {Lopes, Daniel Simões and Parreira, Pedro F. and Mendes, Ana R. and Pires, Vasco M. and Paulo, Soraia F. and Sousa, Carlos and Jorge, Joaquim A.},
	volume = {34},
	copyright = {http://www.springer.com/tdm},
	issn = {0178-2789, 1432-2315},
	url = {http://link.springer.com/10.1007/s00371-017-1448-8},
	doi = {10.1007/s00371-017-1448-8},
	number = {12},
	urldate = {2025-05-23},
	month = dec,
	year = {2018},
	note = {Publisher: Springer Science and Business Media LLC},
	pages = {1713--1723}
}

@article{hyzy_system_2022,
	title = {System {Usability} {Scale} {Benchmarking} for {Digital} {Health} {Apps}: {Meta}-analysis},
	volume = {10},
	copyright = {Unless stated otherwise, all articles are open-access distributed under the terms of the Creative Commons Attribution License (http://creativecommons.org/licenses/by/4.0/), which permits unrestricted use, distribution, and reproduction in any medium, provided the original work ("first published in JMIR mHealth and uHealth...") is properly cited with original URL and bibliographic citation information. The complete bibliographic information, a link to the original publication on http://mhealth.jmir.org/, as well as this copyright and license information must be included.},
	shorttitle = {System {Usability} {Scale} {Benchmarking} for {Digital} {Health} {Apps}},
	url = {https://mhealth.jmir.org/2022/8/e37290},
	doi = {10.2196/37290},
	language = {EN},
	number = {8},
	urldate = {2025-05-23},
	journal = {JMIR mHealth and uHealth},
	author = {Hyzy, Maciej and Bond, Raymond and Mulvenna, Maurice and Bai, Lu and Dix, Alan and Leigh, Simon and Hunt, Sophie},
	month = aug,
	year = {2022},
	note = {Company: JMIR mHealth and uHealth
Distributor: JMIR mHealth and uHealth
Institution: JMIR mHealth and uHealth
Label: JMIR mHealth and uHealth
Publisher: JMIR Publications Inc., Toronto, Canada},
	pages = {e37290},
}

@article{10.5555/2817912.2817913,
author = {Brooke, John},
title = {SUS: a retrospective},
year = {2013},
issue_date = {February 2013},
publisher = {Usability Professionals' Association},
address = {Bloomingdale, IL},
volume = {8},
number = {2},
journal = {J. Usability Studies},
month = feb,
pages = {29–40},
numpages = {12}
}

@misc{ISO9241_110,
  title        = {International Organization for Standardization. (2020). ISO 9241-110:2020 Ergonomics of human-system interaction—Part 110: Interaction principles.},
  url = {https://www.iso.org/standard/75258.html},
  note         = {last accessed: 2025/05/23}
}
\end{document}